# A Multi-technique Elemental and Microphase Analysis in Weathered Arsenic-bearing Granitic Rock Outcrop from Krunkelbach Valley Uranium Deposit, Southern Germany

*Ivan Pidchenko[a,b*], Stephen Bauters[a,b], Irina Sinenko[c], Simone Hempel[d], Lucia Amidani[a,b], Dirk Detollenaere[e,g], Laszlo Vinze[e], Dipanjan Banerjee[f,g], Roelof van Silfhout[h], Stepan Kalmykov[c], Jörg Göttlicher[i], Robert J. Baker[j], Kristina Kvashnina[a,b*]*

[a] The Rossendorf Beamline at ESRF – The European Synchrotron, CS40220, 38043 Grenoble Cedex 9, France

[b] Helmholtz-Zentrum Dresden-Rossendorf (HZDR), Institute of Resource Ecology, PO Box 510119, 01314 Dresden, Germany

[c] Lomonosov Moscow State University, Department of Chemistry, 119991 Moscow, Russia

[d] Technical University Dresden, Institute of Construction Materials, DE-01062 Dresden, Germany

[e] Department of Chemistry, X-ray Imaging and Microspectroscopy Research Group, Ghent University, Ghent, Belgium

[f] Department of Chemistry, KU Leuven, Celestijnenlaan 200F, Box 2404, B-3001 Leuven, Belgium

[g] Dutch-Belgian Beamline (DUBBLE), European Synchrotron Radiation Facility, 71 avenue des Martyrs, CS 40220, 38043 Grenoble Cedex 9, France

[h] The University of Manchester, School of Electrical and Electronic Engineering, Sackville Street, Manchester M13 9PL, United Kingdom

[i] Karlsruhe Institute of Technology, Institute for Photon Science and Synchrotron Radiation (IPS), P.O. 3640, D-76021 Karlsruhe, Germany

[j] School of Chemistry, University of Dublin, Trinity College, Dublin 2, Ireland

ABSTRACT

A synchrotron and laboratory multi-technique investigation has been performed to reveal uranium (U) speciation at an outcropping granitic rock collected from Krunkelbach Valley U deposit area near an abandoned U mine, Black Forest, Southern Germany. The former Krunkelbach mine with 1-2 km surrounding area represents a unique natural analogue site with rich accumulation of secondary U minerals suitable for radionuclide migration studies from a spent nuclear fuel (SNF) repository. Based on a multiple micro-technique analysis using synchrotron-based X-ray fluorescence (µ-XRF), X-ray absorption near edge structure (µ-XANES) spectroscopy and powder X-ray diffraction (µ-XRD) and laboratory-based scanning electron microscopy with energy disperisive X-ray (SEM-EDX) and Raman spectroscopic technique, mixed metazeunerite-metatorbernite, $Cu(UO_2)_2(AsO_4)_{2-x}(PO_4)_x\cdot 8H_2O$, microcrystals are identified on the surface of the rock together with diluted coatings with composition close to cuprosklodowskite, $Cu(UO_2)_2(SiO_3OH)_2\cdot 6H_2O$. In few cavities of the rock well-preserved metatorbernite, $Cu(UO_2)_2(PO_4)_2\cdot 8H_2O$, microcrystals are predominantly identified. The occurrence of the mixed uranyl-arsenate-phosphate and uranyl-silicate mineralization on the surface of same rock indicates on the signatures of different geochemical conditions which took place after the oxidative weathering of the primary U and arsenic (As) ores. Mixed metazeunerite-metatorbernite microcrystals exhibit varying As content depending on the crystal´s part which range from well-preserved to heavily corroded within ~ 200 µm. Uneven alteration is attributed to microheterogeneity of the crystal’s structure, different morphology and chemical composition and varying geochemical conditions at the site following formation of the secondary U mineralization.

The relevance of uranyl minerals to SNF storage and potential role of uranyl-arsenate mineral species on mobilization of U and As into the environment is discussed.

## INTRODUCTION

Uranium (U) is an important trace element and contaminant representing a significant environmental hazard after the mining and ore reprocessing activities.[1-3] U containing natural systems, e.g. ore bodies, former mining sites are often considered as natural analogues for investigations of potential radionuclides release and retardation processes expected in a real spent nuclear fuel (SNF) repository.[4] In this context several mineralogical studies have focused on the alteration and oxidative corrosion processes of a primary U mineral uraninite, $UO_{2+x}$, and SNF's components under ambient and extreme conditions.[5-7] To assess potential risks associated with the long-term storage and possible alteration of SNF, actinide- (An) and lanthanide-containing systems have been intensively investigated in order to draw comparisons with analogue systems in a functional repository.[8-10]

Depending on the local geology and geochemical conditions, the alteration of $UO_{2+x}$ results in the formation of various alteration products. Whilst the richest U mineral families are (oxyhydr)oxides, carbonates and silicates (Si), a smaller number of uranyl minerals are represented by selenates (Se) and arsenates (As), also occurring under oxidizing conditions.[11] Selenium and arsenic are also of environmental concern due to their toxicity (Se, As) and radioactivity ($^{79}$Se).[12-16] The formation of uranyl selenates and arsenates are mainly associated with oxidation processes of sulfide (S) minerals and acidification of groundwaters followed by subsequent release of S, Se, As and other trace elements along with U from associated mineralization. Thus U, S, As and Se traces were identified simultaneously present in the ore material from the former Krunkelbach mine in both unaltered and altered ores.[17] U and As are often associated together in organics rich sediments where U occurs mainly as reduced U(IV) species, mine tailings and in abandoned mining sites after underground flooding activities. This causes additional hazards associated with release of As into water aquifers.[18-24] In cases when reduction conditions prevail, such as at the Rupchechtov site in the Czech Republic, As occurs in the form of arsenopyrite (FeAsS) in tertiary sediments forming layered aggregates with secondary uraninite and arsenopyrite.[19] Mixed uranyl-arsenate-phosphate phases have been identified in the soils from abandoned U mine in the UK as a result of many years of the mining activities at the site.[18, 20] Based on the results of these investigations As is assumed to control U mobility by formation of sparingly soluble $Cu(UO_2)_2(AsO_4)_{2-x}(PO_4)_x{\cdot}8H_2O$ solid solution ($K_{sp} = 10^{-49.2}$). Indeed uranyl arsenates form compounds that have a much lower solubility products compared to other uranyl phases, i.e. U (oxyhydr)oxides with U being often incorporated into Fe (oxyhydr)oxides ($K_{sp} = 10^{-37}$–$10^{-44}$),[25] thus limiting U and As release into the environment.[18, 26] The occurrence of mixed $Cu(UO_2)_2(AsO_4)_{2-x}(PO_4)_x{\cdot}8H_2O$ phase with small amount of $PO_4^{3+}$ substituting for $AsO_4^{3+}$ was first discussed by Frondel.[27] Recent studies of U mineralization from U deposit in Cornwall, Southern UK reported mixed phases with more than 20 at.% of P content. Mineralogical and chemical properties of such mixed phases, however, are still ill defined. Therefore, studies on synthetic and natural species from different geological locations are necessary to provide additional information on the degradation properties of these compounds, i.e. dissolution and ion-exchange behavior depending on chemical composition and temperature.

Both synchrotron and laboratory methods are extensively used separately and in a combination for investigations of structural, redox and degradation properties of U minerals.[18, 20, 23, 28-30] The use of combined experimental approach is often preferred due to the intricate U speciation in environmental systems which helps to develop optimal strategies for contaminated site remediation.[31, 32] Synchrotron methods provide robust and fast analysis with deep sample penetration depth for elemental mapping and speciation while laboratory tools give more detailed analysis on sample´s morphology and more detailed speciation.[33-35] For example, the detailed speciation analysis at U contaminated sites in Ohio, at Oak Ridge and at Hanford Site (all in USA) allowed to develop effective engineering campaigns for reducing U content in groundwaters by in-situ sorption/precipitation or by utilizing permeable reactive barriers.[36-38]

In this work we demonstrate how a combination of synchrotron and laboratory tools can be effectively utilized for the analysis of environmental samples without complicated sample preparation procedure. A case study for elemental and microphase speciation on an outcropping granite rock collected near an abandoned uranium mine in Southern Germany is performed. The advantages of micro techniques: µ-XRF, µ-XANES and laboratory µ-Raman spectroscopies for investigation of complex microphase U-mineral assemblages are highlighted.

## MATERIALS AND METHODS

**Sample description**. A sample of ~ 5 × 5 × 10 $mm^3$ was collected from a granitic rock outcrop near Krunkelbach abandoned U mine in the Menzenschwand, Black Forest (Southern Germany). Krunkelbach uranium deposit is a hydrothermal vein-type deposit with late Carboniferous formation age of 295 ± 7 Ma and the age of secondary U mineralization of 300 ± 50 ka referred to Quaternary period.[39] The pilot exploration took place 1960´s by shaft mining and U reserves estimated at 1000 tons of $U_3O_8$ at an average grade of 0.7%.[40] (see geological map in Figure S1).[22] The sample was first analysed with the Carl Zeiss STEMI 2000C stereomicroscope to select suitable part and microcrystals for further investigations. An area containing visible green crystals mixed with goethite needle was subsequent selected for µ-XRF, U $L_3$ edge µ-XANES and µ-PXRD analyses.

**Micro- X-ray fluorescence (µ-XRF) and U $L_3$ edge micro X-ray absorption near edge structure (U $L_3$ edge µ-XANES).** The µ-XRF and U $L_3$ edge µ-XANES measurements were performed at the DUBBLE BM26A beamline of the European Synchrotron Radiation Facility (ESRF).[41] The incident energy was selected using the a double Si(111) crystal monochromator. Rejection of higher harmonics was achieved with two Pt mirrors at an angle of 2 mrad relative to the incident beam. The dedicated micro-focus platform provided an 8 × 8 $\mu m^2$ spot size at the sample position. XRF mappings were recorded at 17,177 eV with a 1 s dwell time and 20 µm step size. U $L_3$ edge µ-XANES spectra were collected on nine different spots on the area of 1.5 × 2.5 mm. Several spectra were measured at each selected spot for each reference sample: metazeunerite (U-As), metatorbernite (U-P) and cuprosklodowskite (U-Si).

**U $L_3$ edge high-energy resolution fluorescence detected X-ray absorption near edge structure (U $L_3$ edge HERFD-XANES).** The U $L_3$ edge HERFD-XANES measurements were performed using Johann-type emission spectrometers installed at Rossendorf (BM20)[42] and BM14 beamlines of the European Synchrotron (ESRF) in Grenoble, France (see SI for details).

**Micro- powder X-ray diffraction (µ-PXRD).** The µ-PXRD patterns were collected at the SUL-X beamline of KIT synchrotron radiation source. Measurements were done in transmission mode with beam size at sample position of about 150 × 150 $\mu m^2$ on grains with a CCD detector (Photonic Science XDI VHR-2 150). The beamline was operated at an energy of 17,000 eV. D values were calibrated with $LaB_6$ (NIST, 660b) (2 Theta values correspond to $\lambda$ = 0.729684 Å after calibration). Measurements were performed under air and room temperature. Data analysis was performed using FIT2D program and DIFFRAC.EVA V4.3 (Bruker).[43]

**Scanning electron microscopy with energy dispersive X-ray (SEM-EDX).** The SEM-EDX investigations were performed at TU Dresden with a QUANTA 250 FEG (FEI) microscope in LowVac mode combined with an EDX-system QUANTAX 400 (Bruker). The software Esprit 2.1 were used to evaluate the EDX data.

**Raman spectroscopy.** Raman measurements were conducted at room temperature using LabRam ARAMIS (Horiba Jobin Yvon) at excitation wavelength of 532 nm (Nd:YAG). The machine was calibrated on a silicon wafer using the first-order Si line at 520.7 $cm^{-1}$. For all measurements a 1800 lines/mm diffraction grating was used with a slit of 100 µm, a hole of 300 µm, and a neutral density filter D 0.3 (50 % transparency), respectively.

## RESULTS AND DISCUSSION

The sample in our study was collected from the granite outcrop from the Krunkelbach Valley uranium deposit in the early 1970´s and stored under ambient conditions in a mineral collection (see geological map on Figure 1). It was originally described as a two mica granite rock, Bärhalde granite, together with quartz, U silicates – soddyite, $(UO_2)_2SiO_4{\cdot}2H_2O$, U (oxyhyrd)oxide – ianthinite, $U^{4+/5+}(UO_2)_5O_7{\cdot}10H_2O$, and Fe (oxyhydr)oxide – goethite, α-FeO(OH), forming a pseudomorph after $U^{4+/5+}(UO_2)_5O_7{\cdot}10H_2O$ on the surface of a rock. A rich secondary U mineralization represent high environmental significance of the location due to possible degradation of these phases and further migration of dangerous contaminants i.e. U, As and Se in the environment.  In this context one of the aims of the study was to find the evidence for alteration of one of potentially hazardous secondary U phases using a combination of several techniques. To do this we have utilised a multitude of spectroscopic techniques that are both laboratory and synchrotron based. We will firstly describe our characterisation efforts, then put these into the context of SNF storage in a geological repository.

**Elemental and microphase analysis.** Attempts to detect needle shape violet ianthinite crystals initially described on the rock were not successful apparently due to oxidative weathering of the ore over the time it was exposed to ambient, oxidizing conditions. Instead, several tiny platy shaped vitreous green crystals were identified under the optical microscope on the surface and in the cavities of the sample. The µ-XRF element mapping distinguishes regions of different sets of elements with varying signal intensities and areas. U and As are identified in concentrated regions which are associated with Cu (Figure 2A, right column: EDX data with Cu-U-As RGB mapping). Less intense regions show occurrence of Cu, Fe, Pb and W. The later three show no correlation with U and As in the intensive signal regions (see left column XRF on Figure 2A) followed by μ-detailed XANES analysis (Figure 2B). The occurrence of Cu, U and As is in agreement with Cu-bearing uranyl arsenate, $Cu[(UO_2)(AsO_4)_2]$, phase corresponding to (meta)zeunerite, $Cu(UO_2)_2(AsO_4)_2 \cdot 8\text{-}12H_2O$, one of the most common U mineralizations occurring in Krunkelbach area.[44] The presence of metazeunerite (see Figures 3A) was further confirmed by μ-PXRD collected on green crystals selected from the surface (Figure 3B). The presence of W, Bi, Pb and minor Ba in the rock indicates a possible occurrence of two relatively rare U species: uranotungstate, $(Fe,Ba,Pb)(UO_2)_2(WO_4)(OH)_4 \cdot 12H_2O$ and walpurgite, $(BiO)_4(UO_2)(AsO_4)_2 \cdot 2H_2O$. Both species are identified in Krunkelbach and Schneeberg hydrothermal U deposits, respectively.[40, 45] In the Schneeberg deposit $(BiO)_4(UO_2)(AsO_4)_2 \cdot 2H_2O$ is described to occur together with metazeunerite $Cu(UO_2)_2(AsO_4)_2 \cdot 8H_2O$ species. Taking into account that XRF is limited to minimum energy of approximately 6 keV due to photon self-absorption in air, the detection of some elements is hindered. Hence the presence of another U-As mineral phase – nielsbohrite, $K(UO_2)_3(AsO_4)(OH)_4 \cdot H_2O$, cannot be excluded.[46] Other microphases found on the rock correspond to quartz and goethite (Figure 3C) and can be also distinguished on Figure 3A. The energy-dispersive X-ray (EDX) analysis of selected goethite microcrystals did not show presence of U assuming its pseudomorphic nature. In previously reported studies U is found incorporated into a goethite as a result of the oxidation, dissolution/reprecipitation events in a U deposit.[17] Similarly, the correlation of U with Fe minerals: $Cu(UO_2)_2(PO_4)_2 \cdot 8H_2O$ and akageneite, β-FeO(OH), was identified in the soils after intensive U mining activities in Southern UK.[18]

**U speciation by μ-XANES.** Based on μ-XRF map, the U $L_3$ edge μ-XANES spectra were recorded on 5 selected spots with high U content (spectra 1-5) and 4 U containing zones exhibiting less intense U signal (spectra 6-9) (see Figure 2B). In order to compare two sets of recorded spectra, 4 spectral features marked as I, II, III and IV within ~100 eV range, starting from the white line (WL) are highlighted. The spectra exhibit distinct signatures in position of the features II, III and IV, as well as in the shape and intensity of feature I. The intensity of the WL also differs and exhibits a higher intensity in the spectra collected from U diluted zones. All spectra recorded for this zone are similar with spectral features which fits well to cuprosklodowskite, $Cu(UO_2)_2(SiO_3OH)_2{\cdot}6H_2O$; this was previously described as one of uranyl silicate minerals occurred in the deposit.[40] The spectra recorded at concentrated U spots exhibit clear differences of the spectral feature I. The comparison with reference minerals $Cu(UO_2)_2(AsO_4)_2{\cdot}8H_2O$ and $Cu(UO_2)_2(PO_4)_2{\cdot}8H_2O$ and good agreement with XRF analysis shows that the analyzed species are related to these two minerals. Both mineral species are reported for the location studied, with $Cu(UO_2)_2(AsO_4)_2{\cdot}8H_2O$ being the most common secondary U mineral in the area.[44] The $Cu(UO_2)_2(AsO_4)_2{\cdot}8H_2O$ and $Cu(UO_2)_2(PO_4)_2{\cdot}8H_2O$ are isostructural compounds with minor differences in crystallographic parameters. This causes minor differences in XANES spectra and allows for the fingerprinting analysis (see also theoretically modelled XANES spectra on Figure S1).

Additionally, the whole rock sample has been analyzed using U $L_3$ edge HERFD-XANES spectroscopy. The technique in general provides much better resolved spectral features, allowing for significantly more detailed analysis of U redox state especially for environmental relevant systems, where U can be stabilized as a mixture of two or three oxidation states, namely U(IV), U(V) and U(VI).[47-49] The collected spectra exhibit minor differences in the spectral features mainly due to a relatively large beam size, ~ 100 × 400 μm, and likely probing simultaneously several U phases (Figure S2).

**Occurrence of $Cu(UO_2)_2(AsO_4)_{2-x}(PO_4)_x{\cdot}8H_2O$ microphases and evidence for their alteration.** The SEM-EDX performed on one of the green crystals selected from the surface of granitic rock reveals mainly Cu, U, As and lower P (Figure 4, Table 1). In order to find signatures from minor uranyl microphases we used Raman spectroscopy. Raman takes advantage of fast qualitative analysis due to the unique positions of the vibrational bends for metals and ligands.[20, 28] Raman analyses for two different green microcrystals selected from the surface and from the cavities of the rock (see Figure 5A) reveals distinct differences in the U speciation for both microcrystals exhibit vibrational bends characteristic for uranyl ($UO_2^{2+}$), arsenate ($AsO_4^{3-}$) and phosphate ($PO_4^{3-}$): 326 $cm^{-1}$ referred to $\nu_2(AsO_4^{3-})$, 404 $cm^{-1}$ and 458 $cm^{-1}$ described for both $\nu_4(AsO_4^{3-})$ and $\nu_4(PO_4^{3-})$ bending modes (Figure 5B).[20] No information on the peak detected at ~ 493 $cm^{-1}$ is found in the literature. The most intense and typical bends arising at 817-823 $cm^{-1}$ originate from the $\nu_1(UO_2^{2+})$ symmetric stretch (Figure 5C). Two less intense bends arising at 892 $cm^{-1}$ correspond to $\nu_3(UO_2^{2+})$ antisymmetric stretchings in metazeunerite (‘**r1**´) and 992 $cm^{-1}$ – to $\nu_3(PO_4^{3-})$ to metatorbernite, respectively.[20, 28] Another weak vibration normally resolved in metatorbernite 900-905 $cm^{-1}$ is attributed to $\nu_3(UO_2^{2+})$ stretching. This bend apparently overlaps with 870-905 $cm^{-1}$ region characteristic for metazeunerite.[28] Metatorbernite exhibits weak frequency at 900-905 $cm^{-1}$ from the antisymmetric P-O stretching which overlaps with more intense As-O bend appearing at 870-905 $cm^{-1}$.[28] Some broadening of the $\nu_1(UO_2^{2+})$ stretch of both spectra might be a result of the overlapping with $\nu_1(AsO_4^{3-})$ appearing at 815 $cm^{-1}$.[26] Analysis of the Raman spectra for four natural metazeunerite species collected from different locations summarized in RRUFF database gives average $\nu_1(UO_2^{2+})$ = 815 $cm^{-1}$,[50] which agrees with our value, 817 $cm^{-1}$. Some shifts of the frequencies might be attributed to a presence of $AsO_4^{3-}$ and/or other fractions in each species. The evidence that some $AsO_4^{3-}$ might be present in ´**r2**´ is supported by a presence of a feature distinguishable at 825-830 $cm^{-1}$ and by an additional bend at 992 $cm^{-1}$ characteristic explicitly for the $PO_4^{3-}$ group. Spectrum analysis resolved two peaks at 817 $cm^{-1}$ and 827 $cm^{-1}$ which agree well with $\nu_1(UO_2^{2+})$ values for $Cu(UO_2)_2(AsO_4)_2 \cdot 8H_2O$ and $Cu(UO_2)_2(PO_4)_2 \cdot 8H_2O$, respectively (Figure S3).[28, 51]

SEM-EDX analysis of a ~ 250 µm green crystal reveals two discrete parts (Figure 4), a well-preserved part with clearly defined pyramidal part with terminated top plane, [001], and heavily corroded lower part without distinguishable shape and varying chemical composition (Table 1). Dark grey parts in the SEM image belong to goethite debris as well as lighter parts from quartz. EDX analysis of four different crystal's parts shows the highest variation in As and P contents around part ′4′ (Table 1). Part ′1′ belongs to a well-preserved crystal part while parts ′2-3-4′ presumably belong to the overgrown layer with heavily corroded parts ′3′ and ′4′. The latter is associated with a decrease of U and P and some increase of the As content. The most significant variation in elemental composition is found for Cu, U, As and P in crystal's parts ′2′and ′4′. The decrease in U and P content is associated with some increase of As and Cu content in part '4'. The elemental composition of this part is close to the theoretical composition for metazeunerite (Table S1). Degradation processes on a microscale level might be primarily associated with crystal's cleavage and varying elemental composition (see Figure S4 and Table S2). This would result in a heterogenous phase alteration with enhanced dissolution followed by re-precipitation processes on a cleavage planes, retention and/or removal of As and other elements.[40, 52] Following this suggestion, the increase of As in part '4′ might be a result of the local phase dissolution followed by formation of hillocks around altered part enriched with As and Cu. Different geochemical behavior of the corresponding As and P uranyl phases can be attributed to their solubilities: ($K_{sp} = 10^{-49.2}$)[26] for $Cu(UO_2)_2(AsO_4)_2 \cdot 8H_2O$ compared to more soluble ($K_{sp} = 10^{-28.0}$) $Cu(UO_2)_2(PO_4)_2 \cdot 8H_2O$. Thus, the alteration of $Cu(UO_2)_2(AsO_4)_{2-x}(PO_4)_x \cdot 8H_2O$ can be attributed to several factors including the heterogeneity in the crystal's microstructure, morphology, varying chemical composition as well as different geochemical conditions at the site. Highly heterogeneous alteration of $Cu(UO_2)_2(AsO_4)_{2-x}(PO_4)_x \cdot 8H_2O$ can be also related to different chemical stability of the crystal's edges. Similarly, higher stability of certain crystal edges in various U (oxyhydr)oxides was found and explained by low bond-valence deficiency and hence lowest interaction with solution species and the highest stability.[53, 54]

**Environmental impact of the secondary U minerals and their relevance to the SNF storage in geological repositories.** The Krunkelbach deposit represents a potentially rich source of As originating mainly from the (meta)zeunerite reported as the main secondary U mineral species occurring at the site with several individual U-As species described elsewhere (see Figure 6 and Table 2).[22] Formation of $Cu(UO_2)_2(AsO_4)_2 \cdot 8H_2O$ and mixed $Cu(UO_2)_2(AsO_4)_{2-x}(PO_4)_x \cdot 8H_2O$ phases is likely to take place after oxidative leaching of $UO_{2+x}$, FeAsS and $CuFeS_{1-2}$, and $Cu_3AsS_4$, $Cu_{12}(Zn,Fe)_2As_4S_{13}$ as potential source of copper. Phosphorus is released from the host rocks once slightly acidic or close to neutral conditions prevail ($pH \leq 7$).[22, 45] Uranyl silicates are precipitated earlier from Si and U(VI) dissolved in groundwaters under slightly alkaline conditions resulting in a complex U mineralization (see TOC Figure)[5, 31, 45] The knowledge of degradation properties for these environmentally relevant compounds is scarce however. The correlation between U and As assumes that the complexation would strongly affect the geochemistry of these two elements. The stability of uranyl-arsenates can be understood in terms paragenesis of U minerals where U-As together with structurally identical phosphate U form complexes exhibiting extremely low solubility and thus, high stability to groundwaters.[55] Owing to their low solubilities U-P and U-As are considered important species controlling U speciation in the near-surface environment as well as U mobility in natural systems, including different type groundwaters.[56, 57] While U-P are one of the most widespread and abundant species e.g. metatorbernite, the data about U-As species is to

some extent limited. Additional limitations in assessing the geochemical behavior of U-As systems explained by a lack of reliable thermodynamic data for pH > 6.[58]

The U concentration in groundwaters in some of the deposit´s sections has been analyzed to range from ppb to a few ppm U level. The sharp increase in U concentration is associated with the oxidation of the uraninite after the intrusion of oxygenated groundwater.[17] The analysis of U distribution in tap and groundwater shows high median U contents in tap water at 0.76 µg/L for the region of Baden-Württemberg (Southern Germany).[59] Tap water in the southern part of the region however is not systematically monitored for U content. Elevated As content is reported in regions of Bad Herrenalb and Baden Baden where thermal and groundwaters mobilize As from sediments and minerals.[60] Region near to Baden Baden is known as rich deposit of U-As mineralization, i.e. metakirchheimerite, $Co(UO_2)_2(AsO_4)_2 \cdot 8H_2O$, a potential source of both U and As in the groundwaters.[44]

The Krunkelbach deposit has exceptionally rich U mineralogy with more than 40 secondary uranyl mineral species described at a relatively small area of the 240 m deep mine with surrounding location (see Table 2, Figure 6). The species include $UO_{2+x}$, uranyl peroxide studtite, $[(UO_2)(O_2)(H_2O)_2] \cdot H_2O$, and $U^{4+/5+}(UO_2)_5O_7 \cdot 10H_2O$.[61, 62] The latter is considered as an important intermediate species in the paragenesis of U minerals and proposed as a potential mineral phase capable of hosting IV-V valent An, i.e. Pu(IV) and Np(V).[62] The oxidation of uraninite is related to penetration oxygenated water from the water-bearing fractures resulting in formation of the numerous secondary uranyl species. The oxidation processes are estimated to begin 250-350 ky ago continuing up to date causing a loss of up to 10 % of initial U inventory.[17, 22] The location may be therefore considered as a potential ´natural analogue´ site for an operational SNF repository for which safety requirements imply safe storage of the SNF material for more than 100 ky. The detailed knowledge of the geological history, geochemistry and degradation properties of secondary uranyl phases are therefore of crucial significance for the assessment of suitability of the geological sites for long-term SNF storage. The mineralogy of SNF will ultimately determine its durability to self-irradiation effects, chemical corrosion with subsequent release of the radionuclides. Under specific geochemical conditions uranyl phases might serve as solubility controls restricting U migration even when present in highly mobile U(VI) form.[18, 31] During the investigation of the Krunkelbach Valley uranium deposit both unaltered and altered ores analyzed for the As content showed minor release from initially estimated ~ 1200 ppm As content in unaltered rock.[17] One the other hand, much more soluble uranyl phosphate species are reported to restrict U removal by formation of several earth-alkaline uranyl phosphates, i.e. $Ba(UO_2)_2(PO_4)_2{\cdot}8H_2O$ due to higher phosphorus mobility released after oxidative weathering from the related rocks.[22] A minor loss of P and As is reported in the study with the highest release is detected for U owing to its oxidative leaching as a geochemically mobile U(VI) species with.[22] The retardation of the mobilized U has been identified on clay colloids and Ba-phosphate minerals and through precipitation of individual U(VI) mineral species.

More generally, depending on specific geochemical conditions U(VI) minerals can be important species controlling An mobility at U ore reprocessing and mining sites. Uranyl silicates, Ca-uranophane, $Ca(UO_2)_2(SiO_3OH)_2{\cdot}5H_2O$, and $Cu(UO_2)_2(SiO_3OH)_2{\cdot}6H_2O$ and metatorbernite group of minerals are reported to control U speciation at the contaminated Hanford and Oak Ridge sites[36, 63, 64] Similarly, Ca-uranophane and haiweeite, $Ca(UO_2)_2Si_5O_{12}(OH)_2{\cdot}3H_2O$, were found to determine U mobility at the Forsmark, a proposed host for radioactive waste repositories in Sweden.[65] In another U hydrothermal type deposit located in Southern France uranium was found in weathered waste rocks to occur as uranyl phosphate comparable to autunite, $Ca(UO_2)_2(PO_4)_2{\cdot}10\text{-}12H_2O$, linked with monodentate $PO_4^{3-}$ and U(VI) species immobilized on clay minerals.[29] Different U(VI) species, uranocircite, $Ba(UO_2)_2(PO_4)_2{\cdot}8H_2O$, and $Cu(UO_2)_2(AsO_4)_2{\cdot}8H_2O$, dominate in the area of Krunkelbach U deposit resulting after high Ba (higher Ba/Ca ratio) and As contents in the groundwaters and their preferential fixation on altered $UO_{2+x}$.[17, 22] Further oxidative dissolution of $UO_{2+x}$ from the microcavities and surface of the rock favours the release of these elements as well as P, W, Pb, Si and Fe into the environment.[66]

To conclude, in this study we demonstrate how a combination of synchrotron and laboratory techniques can be utilized for a rapid mineralogical analysis of weathered granitic rock without complicated sample preparation and treatment procedures. Based on this analytical approach a multiphase uranium mineralization including $Cu(UO_2)_2(SiO_3OH)_2{\cdot}6H_2O$ coatings and mixed $Cu(UO_2)_2(AsO_4)_{2-x}(PO_4)_x{\cdot}8H_2O$ microcrystals is found. We show the evidence for the microscale chemical and morphological heterogeneities of $Cu(UO_2)_2(AsO_4)_{2-x}(PO_4)_x{\cdot}8H_2O$ phase. The microphase collected from the surface of a rock exhibits highly and unevenly altered morphology with higher As/P ratio, while crystals collected from the cavities of the rock are well-preserved and show mainly $Cu(UO_2)_2(PO_4)_2{\cdot}8H_2O$ phase. These structural heterogeneities and degree of phase alteration can be attributed to local geochemical conditions and weathering time. In the recent study $Cu(UO_2)_2(AsO_4)_{2-x}(PO_4)_x{\cdot}8H_2O$ phases are identified in the soils at abandoned U mine and attributed to intensive U mining activities.[18] In this context the stability of the secondary phases and estimation of their long-term behaviour becomes significant for predicting mobilization of radionuclides. Additional research should be focused on investigation of the thermodynamic and degradation properties of the mixed U-As/U-P phase,[67] analysis of U-As mineralization especially in natural soil systems and linking their occurrence to local geology and geochemical conditions. Geological sites where U-As/U-P phases might occur include abandoned U mines, geological formations considered for the storage of SNF around hydrothermal U deposits of the Orogenic belt in Western and Central Europe including Southern UK.[40]

FIGURES

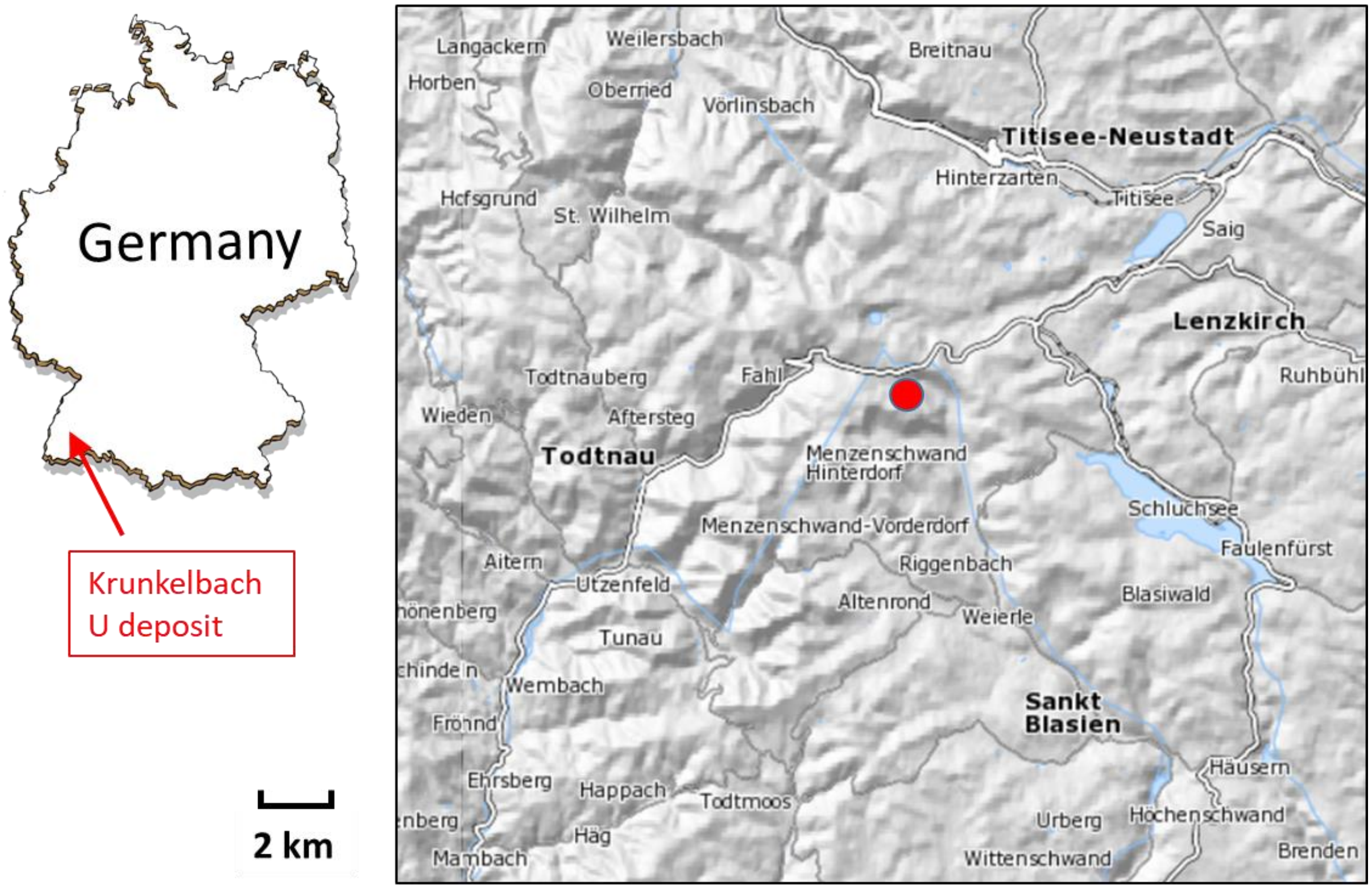

**Figure 1**. Location of the Krunkelbach Valley U deposit. A relief map of the site is retrieved from database of Geological Survey of Germany (maps.lgrb-bw.de).

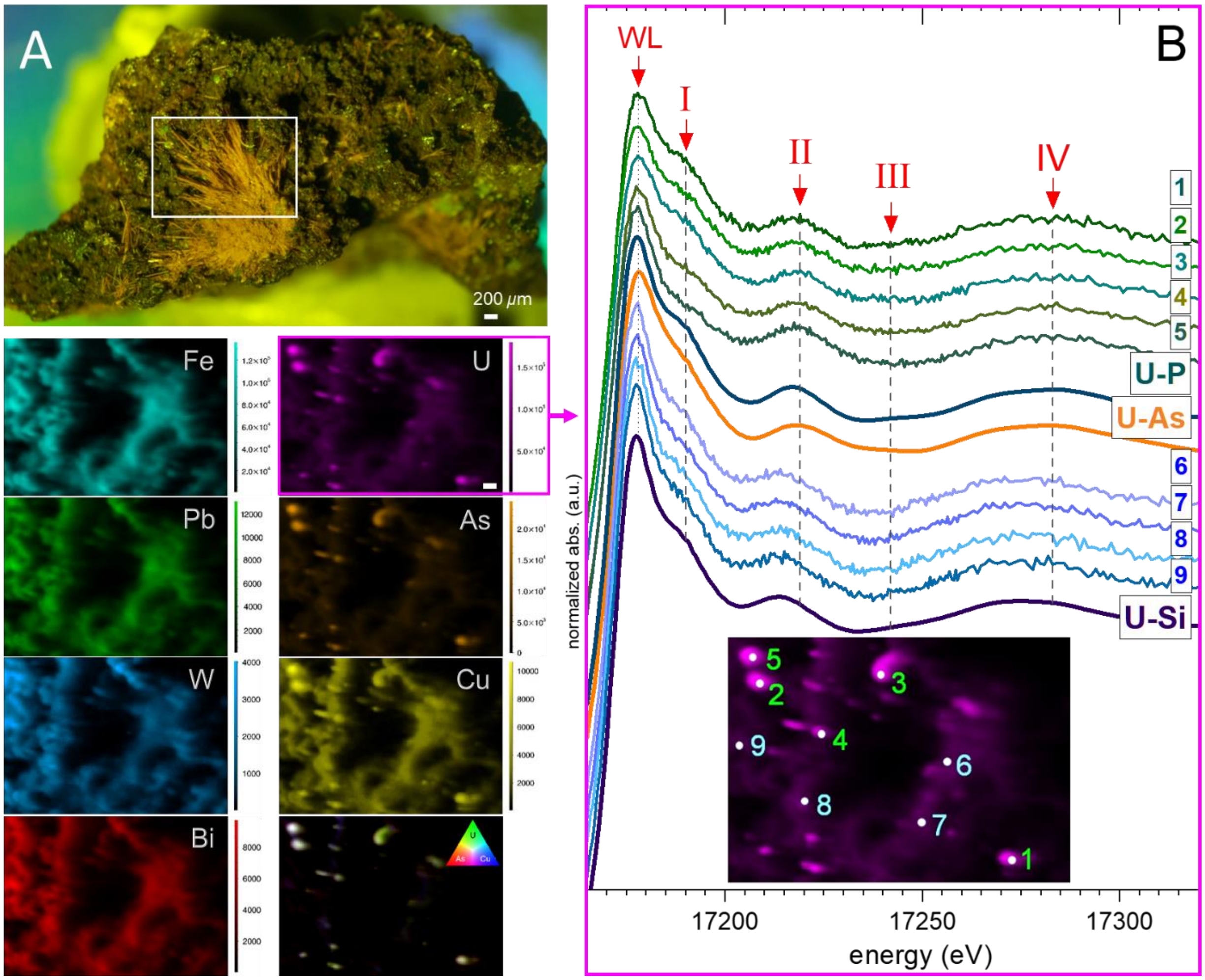

**Figure 2**. Photograph of the rock sample with outlined region (2.5 × 1.5 $mm^2$) of analysis, μ-XRF based element mapping (left column: Fe, Pb, W, Bi; right column U, As, Cu and RGB map for U, As, Cu), scale bar shown at 200 μm (A); U $L_3$ edge μ-XANES spectra of nine spots selected from U μ-XRF map, spectra of metatorbernite (U-P), metazeunerite (U-As) and cuprosklodowskite (U-Si) reference samples (B).

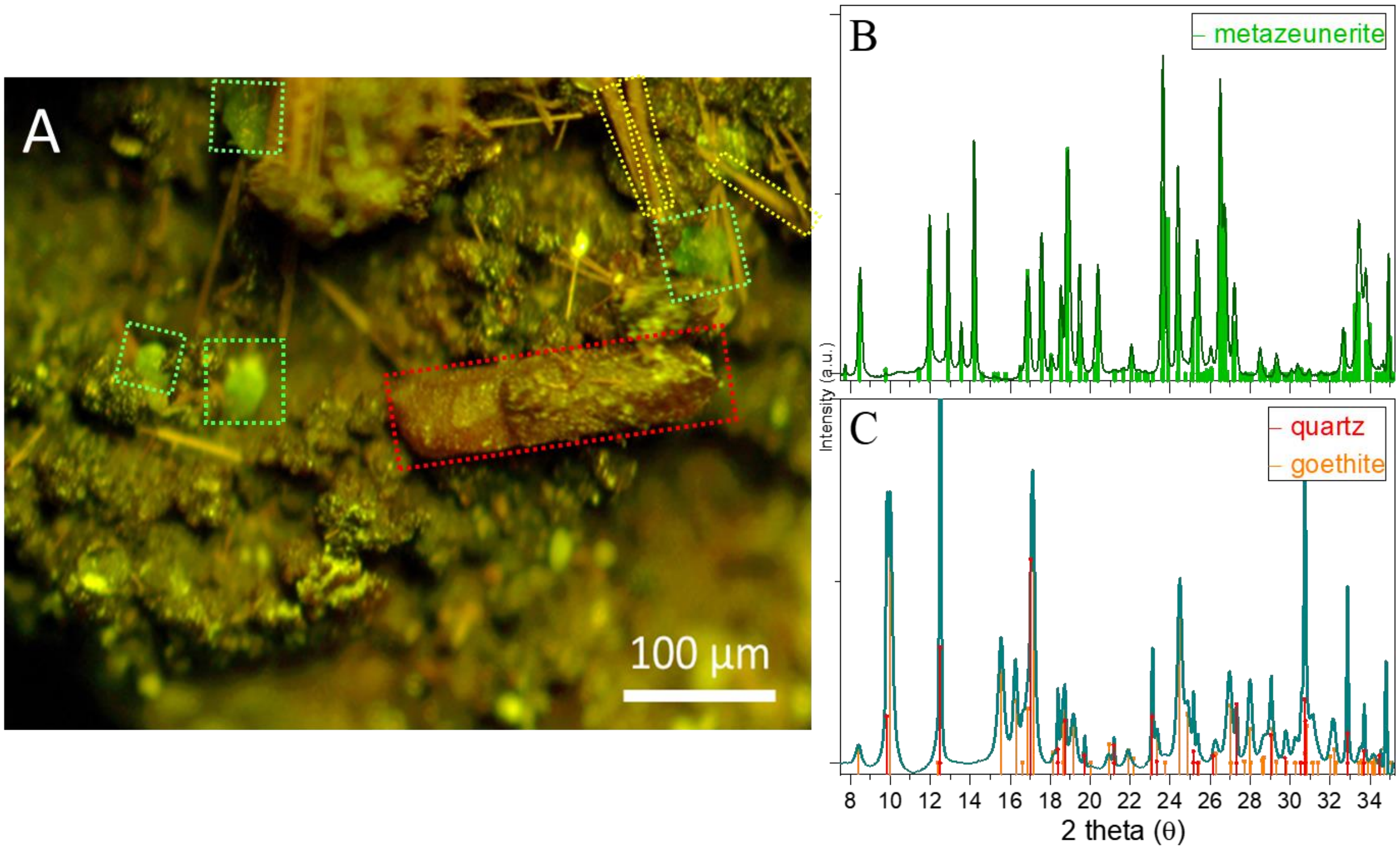

**Figure 3**. Microphotograph of the surface: emerald-green crystals of metazeunerite, dark-brown quartz and light-brown, needle-shaped goethite crystals (A); µ-PXRD patterns of green crystals selected from the surface of the granite outcrop and database metazeunerite (ICDD 40148463)[68] (B), light- and dark-brown crystals correspond to goethite (ICDD 290713) and quartz (ICDD 0898935), respectively (C).

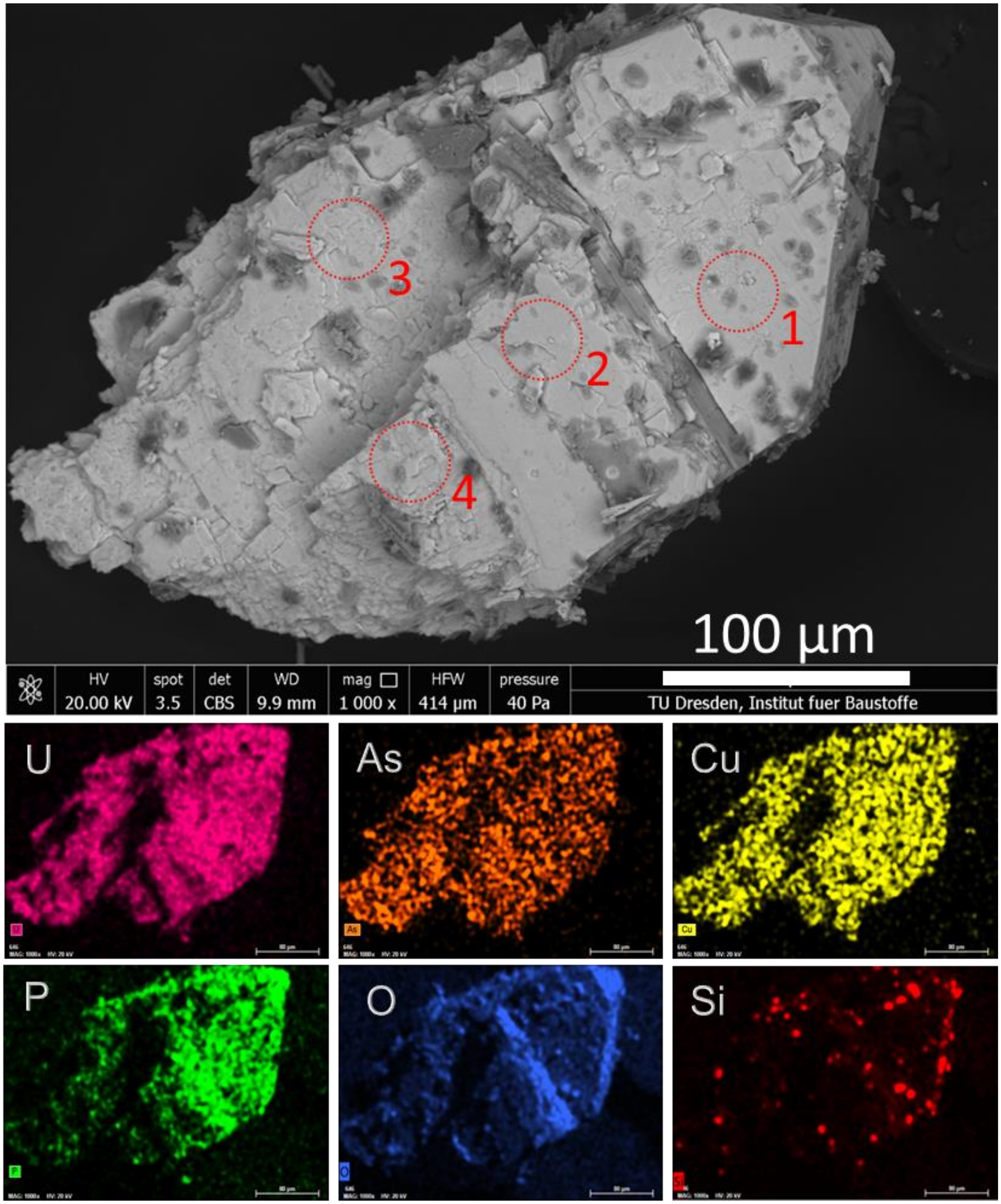

**Figure 4**. SEM image of a green crystal selected from the surface of the granite rock and EDX mappings. Dark grey areas on SEM image correspond to Fe from the goethite debris. Red circles indicate the regions where semiquantitative EDX analyses have been performed (Table 1). (EDX sensitivity for U is estimated at 0.1 wt %).

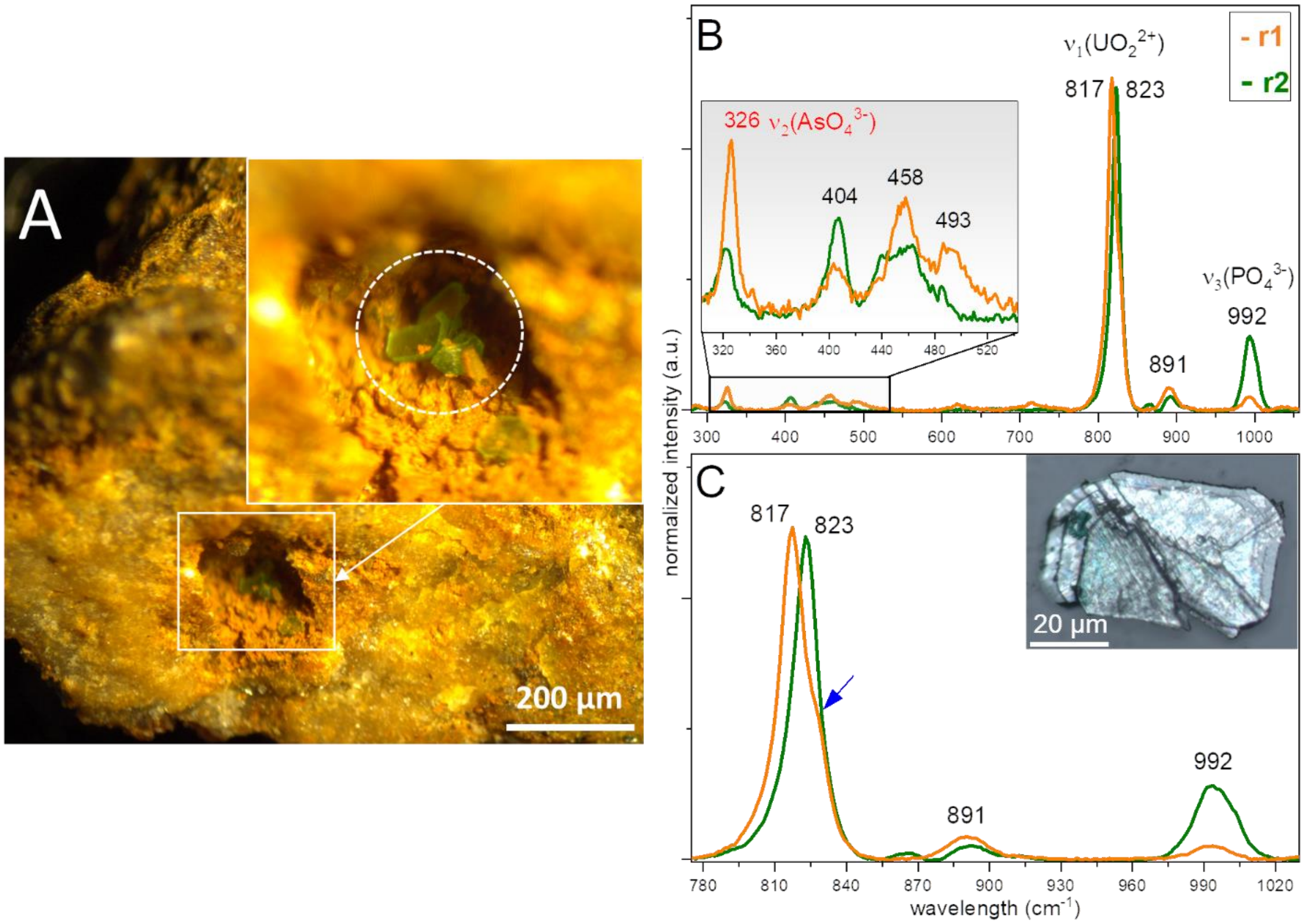

**Figure 5**. Microphotograph of the cavity and flat-shaped vitreous green crystals selected for Raman analysis (A); Raman spectra of two green crystal selected from the surface (´**r1**´, analyzed by µ-PXRD, metazeunerite) and from the cavity (´**r2**´) of the rock with zoomed 290-550 cm$^{-1}$ region (B); Raman spectra for 780-1020 cm$^{-1}$ region. Image of the ~ 50 µm size flat crystal done with Raman spectrometer from ´**r2**´ sample. Blue arrow indicating the spectral feature in ´**r1**´ referring to possible correlation with the metatorbernite phase (C).

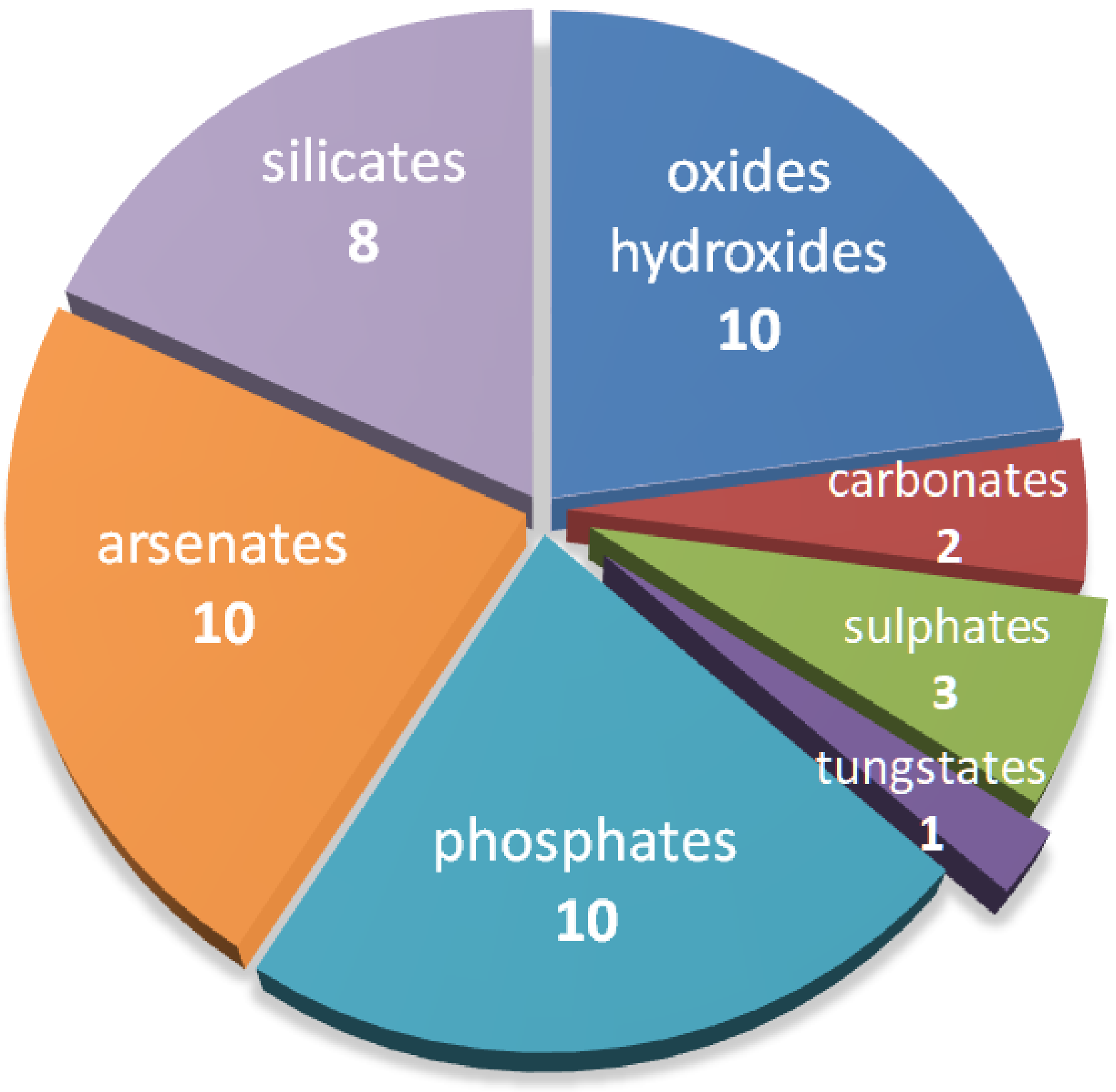

**Figure 6**. Distribution of U minerals by group and number of species identified in Krunkelbach mine (see Table 2).

TABLES

**Table 1**. Concentration of the elements determined from EDX analysis of the $Cu(UO_2)_2(AsO_4)_{2-x}(PO_4)_x{\cdot}8H_2O$ microcrystal (values are given in wt %, deviation ± 1σ).

| Analyzed part | Cu | U | As | P | O |
|---|---|---|---|---|---|
| 1 | 8.9 ± 0.6 | 52.3 ± 1.5 | 11.8 ± 1.0 | 2.8 ± 0.2 | 24.2 ± 4.7 |
| 2 | 7.5 ± 0.7 | 55.8 ± 2.3 | 7.8 ± 0.6 | 2.1 ± 0.3 | 26.8 ± 7.2 |
| 3 | 9.3 ± 0.6 | 55.9 ± 1.9 | 10.9 ± 1.1 | 0.9 ± 0.1 | 22.9 ± 5.4 |
| 4 | 11.5 ± 0.6 | 49.2 ± 1.7 | 14.1 ± 0.8 | 0.8 ± 0.1 | 24.3 ± 4.2 |

**Table 2.** Uranium minerals identified in Krunkelbach mine. [40, 46]

| *Mineral group* | *Chemical formula* |
|---|---|
| *Oxides and hydroxides* | |
| Uraninite/Pitchblende | $UO_{2+x}$ |
| Ianthinite | $U^{4+/5+}(UO_2)_5O_7\cdot10H_2O$ |
| Billietite | $Ba(UO_2)_6O_4(OH)_6\cdot8H_2O$ |
| Wölsendorfite | $(Pb,Ca)U_2O_7\cdot2H_2O$ |
| Schoepite | $(UO_2)_8O_2(OH)_{12}\cdot12H_2O$ |
| Metaschoepite | $(UO_2)_8O_2(OH)_{12}\cdot10H_2O$ |
| Vandendriesscheite | $PbU_2O_7\cdot12H_2O$ |
| Curite | $Pb_3(UO_2)_8O_8(OH)_6\cdot3H_2O$ |
| Clarkeite | $(Na,Ca,Pb)(UO_2)O(OH)\cdot H_2O$ |
| Studtite | $[(UO_2)(O_2)(H_2O)_2]\cdot H_2O$ |
| *Carbonates* | |
| Rutherfordite | $UO_2CO_3$ |
| Joliotite | $(UO_2)CO_3\cdot nH_2O$ |
| *Sulfates* | |
| Zippeite | $K_3(UO_2)_4(SO_4)_2O_3(OH)\cdot3H_2O$ |
| Uranopilite | $(UO_2)_6(SO_4)O_2(OH)_6\cdot14H_2O$ |
| Johannite | $Cu(UO_2)_2(SO_4)_2(OH)_2\cdot8H_2O$ |
| *Tungstates* | |
| Uranotungstate | $(Fe,Ba,Pb)(UO_2)_2(WO_4)(OH)_4\cdot12H_2O$ |
| *Phosphates* | |
| Torbernite | $Cu(UO_2)_2(PO_4)_2\cdot8\text{-}12H_2O$ |
| Metatorbernite | $Cu(UO_2)_2(PO_4)_2\cdot8H_2O$ |
| Autunite | $Ca(UO_2)_2(PO_4)_2\cdot11H_2O$ |
| Metaautunite | $Ca(UO_2)_2(PO_4)_2\cdot6\text{-}8H_2O$ |
| Saleeite | $Mg(UO_2)_2(PO_4)_2\cdot8H_2O$ |
| Uranocircite | $Ba(UO_2)_2(PO_4)_2\cdot8\text{-}12H_2O$ |
| Metauranocircite II | $Ba(UO_2)_2(PO_4)_2\cdot6H_2O$ |
| Bassetite | $Fe^{2+}(UO_2)_2(PO_4)_2\cdot8H_2O$ |
| Bergenite | $Ca_2Ba_4(UO_2)_9(PO_4)_6O_6\cdot16H_2O$ |
| Phosphuranylite | $(H_3O)_3KCa(UO_2)_7(PO_4)_4O_4\cdot8H_2O$ |
| *Arsenates* | |
| Heinrichite | $Ba(UO_2)_2(AsO_4)_2\cdot10H_2O$ |
| Metaheinrichite | $Ba(UO_2)_2(AsO_4)_2\cdot8H_2O$ |
| Zeunerite | $Cu(UO_2)_2(AsO_4)_2\cdot12H_2O$ |
| Metazeunerite | $Cu(UO_2)_2(AsO_4)_2\cdot8\text{-}12H_2O$ |
| Novacekite | $Mg(UO_2)_2(AsO_4)_2\cdot8\text{-}12H_2O$ |
| Abernathyite | $K(UO_2)(AsO_4)\cdot3H_2O$ |
| As-uranospathite | $(F,Cl)_{0.5}(UO_2)_2(AsO_4)_2\cdot20H_2O$ |
| Kahlerite | $Fe(UO_2)_2(AsO_4)_2\cdot12H_2O$ |
| Nielsbohrite | $K(UO_2)_3(AsO_4)(OH)_4\cdot H_2O$ |
| Arsenuranylite | $Ca(UO_2)_4(AsO_4)_2(OH)_4\cdot6H_2O$ |

| *Silicates* | |
|---|---|
| Coffinite | $USiO_4$ |
| Uranosilite | $UO_3 \cdot 7SiO_2$ |
| Soddyite | $(UO_2)_2SiO_4 \cdot 2H_2O$ |
| Uranophane | $Ca(UO_2)_2(SiO_3OH)_2 \cdot 5H_2O$ |
| β-uranophane | $Ca(UO_2)_2(SiO_3OH)_2 \cdot nH_2O$ |
| Ba-uranophane | $Ca(UO_2)_2(SiO_3OH)_2 \cdot nH_2O$ |
| Cuprosklodowskite | $Cu(UO_2)_2(SiO_3OH]_2 \cdot 6H_2O$ |
| Kasolite | $Pb(UO_2)(SiO_4) \cdot H_2O$ |

## ASSOCIATED CONTENT

**Supporting information.** U $L_3$ edge XANES theoretical calculations, details on U $L_3$ edge HERFD-XANES measurement and U $L_3$ edge HERFD-XANES spectra collected from a whole rock, deconvolution of the Raman spectrum, SEM-EDX analysis for some selected microcrystals.

## ACKNOWLEDGEMENTS

This work was supported by European Research Council (ERC) Starting Grant № 759696. I.P. is grateful to Steffen Möckel (Alpha Geophysik GmbH) for providing sample.

## REFERENCES

1. Hähne, R.; Altmann, G. *Principles and Results of Twenty Years of Block-Leaching of Uranium Ores by Wismut GmbH, Germany.* ; IAEA-TECDOC-720, IAEA [external link], Wien, 1993; pp 43-54.

2. Florea, N.; Duliu Octavian, G., Rehabilitation of the Barzava Uranium Mine Tailings. *Journal of Hazardous, Toxic, and Radioactive Waste* **2013,** *17*, (3), 230-236.

3. Abdelouas, A., Uranium Mill Tailings: Geochemistry, Mineralogy, and Environmental Impact. *Elements* **2006,** *2*, (6), 335-341.

4. Petit, J.-C., Natural analogues for the design and performance assessment of radioactive waste forms: A review. *Journal of Geochemical Exploration* **1992,** *46*, (1), 1-33.

5. Baker, R. J., Uranium minerals and their relevance to long term storage of nuclear fuels. *Coordination Chemistry Reviews* **2014,** *266-267*, 123-136.

6. Burns, P. C.; Ewing, R. C.; Navrotsky, A., Nuclear Fuel in a Reactor Accident. *Science* **2012,** *335*, (6073), 1184.

7. Ewing, R. C., Long-term storage of spent nuclear fuel. *Nature Materials* **2015,** *14*, 252.

8. Burakov, B. E.; Ojovan, M. I.; Lee, W. E., *Crystalline Materials for Actinide Immobilisation*. Imperial College Press: 2010; Vol. Volume 1, p 216.

9. Laraia, M., 8 - Remediation of radioactively contaminated sites and management of the resulting waste. In *Radioactive Waste Management and Contaminated Site Clean-Up*, Lee, W. E.; Ojovan, M. I.; Jantzen, C. M., Eds. Woodhead Publishing: 2013; pp 301-326.

10. Miller, W.; Alexander, R.; Chapman, N.; McKinley, J. C.; Smellie, J. A. T., *Geological Disposal of Radioactive Wastes and Natural Analogues*. Elsevier Science: 2000.

11. Burns, P. C., U(VI) Minerals and Inorganic Compounds: Insights Into an Expanded Structural Hierarchy of Crystal Structures. *The Canadian Mineralogist* **2005,** *43*, (6), 1839-1894.

12. Gamaletsos, P., Geological sources of as in the environment of Greece: A review. *The handbook of environmental chemistry* **2016,** *40*, 77-113.

13. Katsoyiannis, I. A.; Hug, S. J.; Ammann, A.; Zikoudi, A.; Hatziliontos, C., Arsenic speciation and uranium concentrations in drinking water supply wells in Northern Greece:

Correlations with redox indicative parameters and implications for groundwater treatment. *Science of The Total Environment* **2007,** *383*, (1), 128-140.

14. Wang, Y.; Le Pape, P.; Morin, G.; Asta, M. P.; King, G.; Bártová, B.; Suvorova, E.; Frutschi, M.; Ikogou, M.; Pham, V. H. C.; Vo, P. L.; Herman, F.; Charlet, L.; Bernier-Latmani, R., Arsenic Speciation in Mekong Delta Sediments Depends on Their Depositional Environment. *Environmental Science & Technology* **2018,** *52*, (6), 3431-3439.

15. Curti, E.; Puranen, A.; Grolimund, D.; Jädernas, D.; Sheptyakov, D.; Mesbah, A., Characterization of selenium in $UO_2$ spent nuclear fuel by micro X-ray absorption spectroscopy and its thermodynamic stability. *Environmental Science: Processes & Impacts* **2015,** *17*, (10), 1760-1768.

16. He, Y.; Xiang, Y.; Zhou, Y.; Yang, Y.; Zhang, J.; Huang, H.; Shang, C.; Luo, L.; Gao, J.; Tang, L., Selenium contamination, consequences and remediation techniques in water and soils: A review. *Environmental Research* **2018,** *164*, 288-301.

17. Hofmann, B. A., Geochemical Analogue Study in the Krunkelbach Mine, Menzenschwand, Southern Germany: Geology and Water-Rock Interaction. *MRS Proceedings* **2011,** *127*, 921.

18. Corkhill, C. L.; Crean, D. E.; Bailey, D. J.; Makepeace, C.; Stennett, M. C.; Tappero, R.; Grolimund, D.; Hyatt, N. C., Multi-scale investigation of uranium attenuation by arsenic at an abandoned uranium mine, South Terras. *npj Materials Degradation* **2017,** *1*, (1), 19.

19. Denecke, M. A.; Somogyi, A.; Janssens, K.; Simon, R.; Dardenne, K.; Noseck, U., Microanalysis (micro-XRF, micro-XANES, and micro-XRD) of a Tertiary Sediment Using Microfocused Synchrotron Radiation. *Microscopy and Microanalysis* **2007,** *13*, (3), 165-172.

20. Driscoll, R. J. P.; Wolverson, D.; Mitchels, J. M.; Skelton, J. M.; Parker, S. C.; Molinari, M.; Khan, I.; Geeson, D.; Allen, G. C., A Raman spectroscopic study of uranyl minerals from Cornwall, UK. *RSC Advances* **2014,** *4*, (103), 59137-59149.

21. Essilfie-Dughan, J.; Hendry, M. J.; Warner, J.; Kotzer, T., Solubility Controls of Arsenic, Nickel, and Iron in Uranium Mine Tailings. In *The New Uranium Mining Boom: Challenge and Lessons learned*, Merkel, B.; Schipek, M., Eds. Springer Berlin Heidelberg: Berlin, Heidelberg, 2012; pp 325-334.

22. Hofmann, B. *Genese, Alteration und rezentes Fliess-System der Uranlagerstätte Krunkelbach (Menzenschwand, Südschwarzwald)*; NAGRA Technischer Bericht 88-30.: 1989; p 280.

23. Noseck, U.; Brasser, T.; Suksi, J.; Havlová, V.; Hercik, M.; Denecke, M. A.; Förster, H.-J., Identification of uranium enrichment scenarios by multi-method characterisation of immobile uranium phases. *Physics and Chemistry of the Earth, Parts A/B/C* **2008,** *33*, (14), 969-977.

24. Paul, M.; Metschies, T.; Frenzel, M.; Meyer, J., The Mean Hydraulic Residence Time and Its Use for Assessing the Longevity of Mine Water Pollution from Flooded Underground Mines. In *The New Uranium Mining Boom: Challenge and Lessons learned*, Merkel, B.; Schipek, M., Eds. Springer Berlin Heidelberg: Berlin, Heidelberg, 2012; pp 689-699.

25. Schwertmann, U., Solubility and dissolution of iron oxides. *Plant and Soil* **1991,** *130*, (1), 1-25.

26. Vochten, R.; Goeminne, A., Synthesis, crystallographic data, solubility and electrokinetic properties of meta-zeunerite, meta-kirchheimerite and nickel-uranylarsenate. *Physics and Chemistry of Minerals* **1984,** *11*, (2), 95-100.

27. Frondel, C. *Systematic mineralogy of uranium and thorium*; 1064; 1958.

28. Faulques, E.; Kalashnyk, N.; Massuyeau, F.; Perry, D. L., Spectroscopic markers for uranium(VI) phosphates: a vibronic study. *RSC Advances* **2015,** *5*, (87), 71219-71227.

29. Tayal, A.; Conradson, S. D.; Kanzari, A.; Lahrouch, F.; Descostes, M.; Gerard, M., Uranium speciation in weathered granitic waste rock piles: an XAFS investigation. *RSC Advances* **2019,** *9*, (21), 11762-11773.

30. Thompson, H. A.; Brown, G. E., Jr.; Parks, G. A., XAFS spectroscopic study of uranyl coordination in solids and aqueous solution. *American Mineralogist* **1997,** *82*, (5-6), 483-496.

31. Maher, K.; Bargar, J. R.; Brown, G. E., Environmental Speciation of Actinides. *Inorganic Chemistry* **2013,** *52*, (7), 3510-3532.

32. Newsome, L.; Morris, K.; Lloyd, J. R., The biogeochemistry and bioremediation of uranium and other priority radionuclides. *Chemical Geology* **2014,** *363*, 164-184.

33. Edwards, N. P.; Webb, S. M.; Krest, C. M.; van Campen, D.; Manning, P. L.; Wogelius, R. A.; Bergmann, U., A new synchrotron rapid-scanning X-ray fluorescence (SRS-XRF) imaging station at SSRL beamline 6-2. *Journal of Synchrotron Radiation* **2018,** *25*, (5), 1565-1573.

34. Hall, C.; Barnes, P.; Cockcroft, J. K.; Jacques, S. D. M.; Jupe, A. C.; Turrillas, X.; Hanfland, M.; Häusermann, D., Rapid whole-rock mineral analysis and composition mapping by synchrotron X-ray diffraction. *Analytical Communications* **1996,** *33,* (7), 245-248.

35. Stefaniak, E. A.; Alsecz, A.; Frost, R.; Máthé, Z.; Sajó, I. E.; Török, S.; Worobiec, A.; Van Grieken, R., Combined SEM/EDX and micro-Raman spectroscopy analysis of uranium minerals from a former uranium mine. *Journal of Hazardous Materials* **2009,** *168,* (1), 416-423.

36. McKinley, J. P.; Zachara, J. M.; Liu, C.; Heald, S. C.; Prenitzer, B. I.; Kempshall, B. W., Microscale controls on the fate of contaminant uranium in the vadose zone, Hanford Site, Washington. *Geochimica et Cosmochimica Acta* **2006,** *70,* (8), 1873-1887.

37. Cravotta, C. A.; Watzlaf, G. R., Chapter 2 - Design and Performance of Limestone Drains to Increase pH and Remove Metals from Acidic Mine Drainage. In *Handbook of Groundwater Remediation using Permeable Reactive Barriers*, Naftz, D. L.; Morrison, S. J.; Fuller, C. C.; Davis, J. A., Eds. Academic Press: San Diego, 2003; pp 19-66.

38. Fuller, C. C.; Bargar, J. R.; Davis, J. A.; Piana, M. J., Mechanisms of Uranium Interactions with Hydroxyapatite:  Implications for Groundwater Remediation. *Environmental Science & Technology* **2002,** *36,* (2), 158-165.

39. Hofmann, B.; Eikenberg, J., The Krunkelbach uranium deposit, Schwarzwald, Germany; correlation of radiometric ages (U-Pb, U-Xe-Kr, K-Ar, 230 Th- 234 U). *Economic Geology* **1991,** *86,* (5), 1031-1049.

40. Bültemann, H. W.; Bültemann, W. D. *The Uranium Deposit Krunkelbach in the Southern Black Forest, Federal Republic of Germany*; IAEA-TECDC-361, 1986; pp 247-260.

41. Bauters, S.; Tack, P.; Rudloff-Grund, J. H.; Banerjee, D.; Longo, A.; Vekemans, B.; Bras, W.; Brenker, F. E.; van Silfhout, R.; Vincze, L., Polycapillary Optics Based Confocal Micro X-ray Fluorescence and X-ray Absorption Spectroscopy Setup at The European Synchrotron Radiation Facility Collaborative Research Group Dutch–Belgian Beamline, BM26A. *Analytical Chemistry* **2018,** *90*, (3), 2389-2394.

42. Kvashnina, K. O.; Scheinost, A. C., A Johann-type X-ray emission spectrometer at the Rossendorf beamline. *Journal of Synchrotron Radiation* **2016,** *23*, (3), 836-841.

43. Hammersley, A., FIT2D: a multi-purpose data reduction, analysis and visualization program. *Journal of Applied Crystallography* **2016,** *49*, (2), 646-652.

44. Walenta, K., Beiträge zur Kenntnis seltener Arsenatmineralien unter besonderer Berücksichtigung von Vorkommen des Schwarzwaldes. *Tschermaks mineralogische und petrographische Mitteilungen* **1964,** *9*, (1), 111-174.

45. Krivovichev, S. V.; Plášil, J., Mineralogy and Crystallography of Uranium. In *Uranium - Cradle to Grave,* In: Burns, P. C.; Sigmon, G. E., Eds. The Mineralogical Association of Canada, Winnipeg, MB, 2013; pp 43-49.

46. Walenta, K.; Hatert, F.; Theye, T.; Lissner, F.; Röller, K., Nielsbohrite, a new potassium uranyl arsenate from the uranium deposit of Menzenschwand, southern Black Forest, Germany. *European Journal of Mineralogy* **2009,** *21*, (2), 515-520.

47. Pidchenko, I.; Kvashnina, K. O.; Yokosawa, T.; Finck, N.; Bahl, S.; Schild, D.; Polly, R.; Bohnert, E.; Rossberg, A.; Göttlicher, J.; Dardenne, K.; Rothe, J.; Schäfer, T.; Geckeis, H.;

Vitova, T., Uranium Redox Transformations after U(VI) Coprecipitation with Magnetite Nanoparticles. *Environmental Science & Technology* **2017,** *51*, (4), 2217-2225.

48. Roberts, H. E.; Morris, K.; Law, G. T. W.; Mosselmans, J. F. W.; Bots, P.; Kvashnina, K.; Shaw, S., Uranium(V) Incorporation Mechanisms and Stability in Fe(II)/Fe(III) (oxyhydr)Oxides. *Environmental Science and Technology Letters* **2017,** *4*, (10), 421-426.

49. Rothe, J.; Altmaier, M.; Dagan, R.; Dardenne, K.; Fellhauer, D.; Gaona, X.; González-Robles Corrales, E.; Herm, M.; Kvashnina, O. K.; Metz, V.; Pidchenko, I.; Schild, D.; Vitova, T.; Geckeis, H., Fifteen Years of Radionuclide Research at the KIT Synchrotron Source in the Context of the Nuclear Waste Disposal Safety Case. *Geosciences* **2019,** *9*, (2).

50. Lafuente, B.; Downs, R. T.; Yang, H.; Stone, N., 1. The power of databases: The RRUFF project. In *Highlights in Mineralogical Crystallography*, 2015.

51. Frost, R. L.; Weier, M. L.; Adebajo, M. O., Thermal decomposition of metazeunerite—a high-resolution thermogravimetric and hot-stage Raman spectroscopic study. *Thermochimica Acta* **2004,** *419*, (1), 119-129.

52. Mereiter, K., The crystal structure of walpurgite, $(UO_2)Bi_4O_4(AsO_4)_2 \cdot 2H_2O$. *Tschermaks mineralogische und petrographische Mitteilungen* **1982,** *30*, (2), 129-139.

53. Schindler, M.; Hawthorne, F. C., A bond-valence approach to the uranyl-oxide hydroxy-hydrate minerals: Chemical composition and occurrence. *Canadian Mineralogist* **2004,** *42*, (6), 1601-1627.

54. Schindler, M.; Mandaliev, P.; Hawthorne, F. C.; Putnis, A., Dissolution of uranyl-oxide-hydroxy-hydrate minerals. I. Curite. *Canadian Mineralogist* **2006,** *44*, (2), 415-431.

55. Korzeb, S. L.; Foord, E. E.; Lichte, F. E., The chemical evolution and paragenesis of uranium minerals from the Ruggles and Palermo granitic pegmatites, New Hampshire. *The Canadian Mineralogist* **1997,** *35*, (1), 135-144.

56. Finch, R.; Murakami, T., Systematics and paragenesis of uranium minerals. In *Uranium: Mineralogy, Geochemistry, and the Environment*, 1999; Vol. 38.

57. Murakami, T.; Ohnuki, T.; Isobe, H.; Sato, T., Mobility of uranium during weathering. *American Mineralogist* **1997,** *82*, (9-10), 888-899.

58. Merkel, B. J., Thermodynamic Data Dilemma. In *The New Uranium Mining Boom: Challenge and Lessons learned*, Merkel, B.; Schipek, M., Eds. Springer Berlin Heidelberg: Berlin, Heidelberg, 2012; pp 627-633.

59. Birke, M.; Rauch, U.; Lorenz, H.; Kringel, R., Distribution of uranium in German bottled and tap water. *Journal of Geochemical Exploration* **2010,** *107*, (3), 272-282.

60. Bender, K. Herkunft und Entstehung der Mineral- und Thermalwässer im nördlichen Schwarzwald. Ruprecht-Karls-Universität, Heidelberg, 1995.

61. Belai, N.; Frisch, M.; Ilton, E. S.; Ravel, B.; Cahill, C. L., Pentavalent uranium oxide via reduction of $[UO_2]^{2+}$ under hydrothermal reaction conditions. *Inorganic Chemistry* **2008,** *47*, (21), 10135-10140.

62. Burns, P. C.; Finch, R. J.; Hawthorne, F. C.; Miller, M. L.; Ewing, R. C., The crystal structure of ianthinite, $[U_2^{4+}(UO_2)_4O_6(OH)_4(H_2O)_4](H_2O)_5$: a possible phase for $Pu^{4+}$ incorporation during the oxidation of spent nuclear fuel. *Journal of Nuclear Materials* **1997,** *249*, (2), 199-206.

63. Phillips, D. H.; Watson, D. B.; Roh, Y., Uranium Deposition in a Weathered Fractured Saprolite/Shale. *Environmental Science & Technology* **2007,** *41*, (22), 7653-7660.

64. Stubbs, J. E.; Elbert, D. C.; Veblen, D. R.; Zhu, C., Electron Microbeam Investigation of Uranium-Contaminated Soils from Oak Ridge, TN, USA. *Environmental Science & Technology* **2006,** *40*, (7), 2108-2113.

65. Krall, L.; Sandström, B.; Tullborg, E.-L.; Evins, L. Z., Natural uranium in Forsmark, Sweden: The solid phase. *Applied Geochemistry* **2015,** *59*, 178-188.

66. Finch Robert, J.; Miller Mark, L.; Ewing Rodney, C., Weathering of Natural Uranyl Oxide Hydrates: Schoepite Polytypes and Dehydration Effects. In *Radiochimica Acta*, 1992; Vol. 58-59, p 433.

67. Kulaszewska, J.; Dann, S.; Warwick, P.; Kirk, C., Solid solution formation in the metatorbernite-metazeunerite system ($Cu(UO)_2(PO_4)_x(AsO_4)_{2-x}\cdot nH_2O$) and their stability under conditions of variable temperature. *Philosophical Transactions of the Royal Society A: Mathematical, Physical and Engineering Sciences* **2019,** *377*, (2147).

68. Hennig, C.; Reck, G.; Reich, T.; Roßberg, A.; Kraus, W.; Sieler, J., EXAFS and XRD investigations of zeunerite and meta-zeunerite. In *Zeitschrift für Kristallographie - Crystalline Materials*, 2003; Vol. 218, p 37.

**TOC Figure**

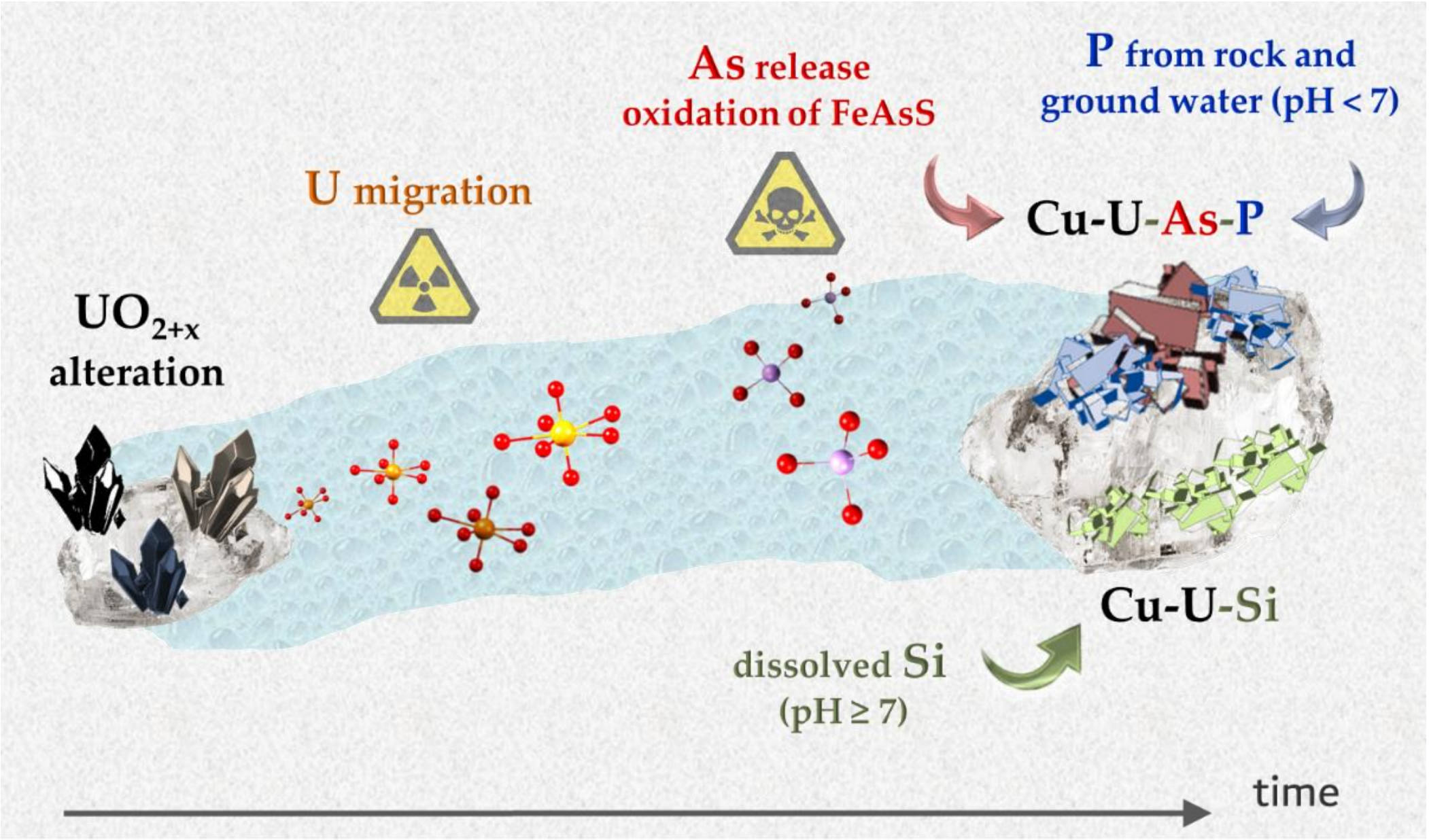